\documentclass[fleqn,usenatbib]{mnras}

\usepackage{newtxtext,newtxmath}

\usepackage[T1]{fontenc}
\usepackage{ae,aecompl}
\usepackage{threeparttablex}
\usepackage{footnote}

\usepackage{graphicx}	
\usepackage{subfig}
\usepackage{longtable}
\usepackage{multirow}
\usepackage{multirow,makecell}
\usepackage{array}
\usepackage{xcolor}
\usepackage{hyperref} 
\usepackage{cleveref}
\usepackage{colortbl}
\newcolumntype{P}[1]{>{\centering\arraybackslash}p{#1}}

\title[Models for SNR populations]{Optical emission-line Luminosity Function models for populations of Supernova Remnants}

\author[Kopsacheili et al.]{
M. Kopsacheili,$^{1, 2}$\thanks{E-mail: mariakop@physics.uoc.gr}
A. Zezas,$^{1,2,3}$
I. Leonidaki$^{1, 2}$
\\
$^{1}$Department of Physics, University of Crete, Heraklion GR-70013, Greece \\
$^{2}$Institute of Astrophysics, FORTH, GR-71110 Heraklion, Greece\\
$^{3}$Harvard-Smithsonian Center for Astrophysics, 60 Garden Street, Cambridge, MA 02138, USA\\
}

\date{Accepted XXX. Received YYY; in original form ZZZ}

\pubyear{2021}

\begin{document}
\label{firstpage}
\pagerange{\pageref{firstpage}--\pageref{lastpage}}
\maketitle

\newcommand{\kms}{km\,s$^{-1}\,$}
\newcommand{\ergs}{erg\,s$^{-1}\,$}
\newcommand{\cmd}{cm$^{-3}\,$}
\newcommand{\hbeta}{H$\rm \beta\,$}
\newcommand{\halpha}{H$\rm \alpha\,$}
\newcommand{\sii}{[\ion{S}{II}]\,}
\newcommand{\ratio}{[\ion{S}{II}]/H$\rm \alpha\,$}
\begin{abstract}
We present a basic model for the calculation of the luminosity distribution of supernova remnant populations. We construct theoretical \halpha and  joint \sii - \halpha luminosity functions for supernova remnants by combining prescriptions from a basic evolution model that provides the shock velocity and radius for SNRs of different age and pre-shock density, with shock excitation models that give the  gas emissivity for shocks of different physical parameters. We assume a flat age distribution, and we explore the effect of different pre-shock density distributions or different magnetic parameters. We find very good agreement between the shape of the model \halpha and the joint \sii - \halpha luminosity functions and those measured from SNR surveys in nearby galaxies.

\end{abstract}




\section{Introduction}
Supernova remnants (SNRs) play significant role in shaping the interstellar medium (ISM; e.g. \citealt{2017ARA&A..55...59N}). They enrich it with metals and they provide large amounts of mechanical energy  which  can play significant role in the evolution of the host galaxy. The propagating shock wave  compresses the ISM and under appropriate conditions it can trigger star formation. Moreover, since core-collapse SNRs are the final product in a massive stars' life, they trace the on-going massive star formation rate (e.g. \citealt{2007PhR...442...38J}; \citealt{1997ARA&A..35..309F}).

\par The study of SNR populations and their physical properties is very important in order to understand their feedback to the ISM and their role in the host galaxy. Systematic studies of Galactic SNRs provide more detailed information about the physical parameters of SNRs and their interaction with their surrounding ISM (\citealt{2020ApJ...898L..51W}; \citealt{2013ApJ...772..134M}; \citealt{2009A&A...499..789B}; \citealt{1985ApJ...292...29F}).  On the other hand, the study of extra-galactic SNRs, gives us the opportunity to examine larger samples of SNRs, in different environments (e.g. different metallicities; \citealt{2019ApJ...875...85L}; \citealt{2013MNRAS.429..189L};\citeyear{2010ApJ...725..842L}; \citealt{2007AJ....133.1361P}; \citealt{1997ApJS..113..333M}; \citealt{1997ApJS..108..261B}). In fact, combining measurements of the emission of SNRs in  diagnostic spectral lines it has been possible to determine their physical parameters (e.g. density or shock velocity) in different galactic environments (e.g. \citealt{2021ApJ...908...80W}; \citealt{2019ApJ...875...85L}; \citealt{2018ApJ...855..140L}; \citealt{2017ApJ...839...83W}; \citealt{2013MNRAS.429..189L}). This way we can study their evolution and understand their interplay with the ISM in conditions representing typical galactic environments.

\par Despite the growing number of known SNRs and SNR populations in other galaxies, we still lack a framework that can describe the SNR populations in terms of their observational characteristics but also in the context of expectations from theoretical models for their evolution. A first step in this direction was made in  \citet{Kopsacheili2021}, who presented a method for the calculation of luminosity functions of SNRs free of selection effects, and introduced the joint \halpha - [\ion{S}{II}] luminosity function (LF), as well as, their excitation function (i.e. the offset of an SNR from the $\rm L_{[S\, II]} = 0.4L_{H\alpha}$ relation). While the \halpha LF provides information on their overall population and energetics, the joint \halpha - [\ion{S}{II}] LF and the excitation function also reflect their interaction with the ISM. The latter in particular, bears the imprint of the shock velocity distribution of the SNRs, in their optical emitting phase.

However, what is missing, is a theoretical framework that can predict the distributions of observable parameters such as the luminosity in continuum bands or emission lines, based on the  distributions of the SNR physical parameters (e.g. density, shock velocity, radius).  Such a framework is useful for exploring selection effects in the observed population, mapping the underlying distribution of the physical parameters  that determine the observed properties of SNRs, or even testing SNR evolution models. One can then use the derived distribution of SNR physical parameters in order to estimate their total luminosity or integrated mechanical energy.

\par  In this work, we present a framework for the calculation of the luminosity function of SNR populations, by combining a basic model for the evolution of SNR physical parameters, with  the shock models from MAPPINGS III \citep{2008ApJS..178...20A}. These models have been used extensively  for the exploration of the physical conditions of extragalactic SNRs. In this work we use them as a predictive tool along with a  baseline model for SNR evolution in the adiabatic and radiative phase in order to derive \halpha and  \halpha - \sii luminosity functions for different assumptions on the pre-shock density distributions or the strength of the magnetic field. We also examine the dependence of the derived luminosity functions on these assumptions. This work is the first step for the construction of population synthesis models of optical SNRs.



\par In $\S$\ref{models}  we describe the basic assumptions of our models and the framework for the calculation of luminosity functions. In $\S$\ref{results} we present the theoretical \halpha and the joint \sii - \halpha luminosity functions derived from this model. In $\S$\ref{discussion} we discuss the results of our model focusing on the effect of different physical parameters and the comparison with observational results. In $\S$ \ref{conlusions} we summarize our results.

\section{Population models} \label{models}
 The evolution of SNRs can be described by four consecutive phases: free expansion, adiabatic (or Sedov-Taylor), radiative, and fade-out phase (e.g. \citealt{1975ApJ...200..698C}; \citealt{1988ApJ...334..252C}). Owing to the energetics and the duration of each of these phases the observed SNRs are mostly either in the adiabatic or the radiative phase. The first is dominated by SNRs emitting in the X-ray band, while SNRs in the latter phase emit predominantly in the optical bands (although there are also optically emitting SNRs in the adiabatic phase and SNRs in the radiative phase emitting in the X-ray band). Since in this work we are interested in the optical emission of SNRs, we focus on the adiabatic and radiative phases to develop a model for the luminosity function of SNR populations.

 \par The transition time between the two phases depends on the density and the metallicity of the surrounding material and the explosion energy. There are more than one definitions for the transition time between Sedov-Taylor and radiative  phase. For example, according to \citet{2011piim.book.....D} the transition happens when 1/3 of the SN energy is already radiated (during the Sedov-Taylor phase). However, for this work we adopt the definition of \citet{1988ApJ...334..252C} according to which, the transition takes place near the shell-formation time $t_{sf}$, when the temperature of the first element of gas becomes zero. In this case $t_{sf} = 3.61\times 10^4\frac{E_{51}^{3/14}}{\zeta_m^{5/14}n^{4/7}\,yr}$, where $E_{51}=E\rm _0(erg)/10^{51}$ is the total energy of the explosion in units of $\rm 10^{51}\,erg$, $n(\rm cm^{-3})$ is the pre-shock density, and $\zeta_m$ is the metallicity factor which is 1 for solar abundances (e.g. \citealt{1982ASIC...90..433M}). 
 
 Hence, the transition time $t_{tr}$ is given by the relation:
 \begin{equation} 
\begin{aligned}
 t_{tr}=\frac{t_{sf}}{e}\, yr
 \label{eq:tr}
\end{aligned}
\end{equation}
where $e$ is the base of the natural logarithm.

The radius and shock velocity in each of these two phases as a function of the pre-shock density $n$ and the SNR age $t$ are given by:

\begin{equation} 
\begin{aligned}
& R_{ST}(n, t) = (\xi 10^{51}E_{51})^{1/5}\rho^{-1/5}(3.3\times 10^{7}t(yrs))^{2/5}\, cm\\
& v_{ST} =  0.4(\xi 10^{51}E_{51})^{1/5}\rho^{-1/5}(3.3\times 10^{7}t(yrs))^{-3/5} \, km\,s^{-1}
\label{eq:R_sedov}\\
\end{aligned}
\end{equation}
for the adiabatic phase, and:
\begin{equation} 
\begin{aligned}
& R_{rad}(n, t) = R_{tr}(n) (\frac{4t}{3t_{tr}(n)} - \frac{1}{3})^{3/10}\,cm\\
& v_{rad}(n, t) = v_{tr}(n) (\frac{4t}{3t_{tr}(n)} - \frac{1}{3})^{-7/10} \, km\,s^{-1}
\label{eq:snowplow}\\
\end{aligned}
\end{equation}
for the radiative phase,\\
where $R_{tr}(n) = 14\frac{E_{51}^{2/7}}{n^{3/7}\zeta_m^{1/7}} 3.1\times 10^{18}\, cm$ and $v_{tr}(n) = 413n^{1/7}\zeta_m^{3/14}E_{51}^{1/14}\, km\,s^{-1}$ are the radius and velocity at the beginning of the radiative phase. In the above relations $\xi = 2.026$ a numerical constant (\citealt{1988ApJ...334..252C}), $\zeta_m$ is the metallicity factor as mentioned earlier, $\rho = \mu_{H}\, n\, m_{H} = 2.3\times 10^{-24}\, gr\, cm^{-3}$ the total mass density, $\mu_H$ the
mean mass per hydrogen nucleus, and $m_H$ the hydrogen-atom mass.
Here we note that the detailed exploration of the evolutionary phase of SNRs is beyond the scope of this paper, and we refer to the two phases (Sedov-Taylor and radiative) only in order to use the appropriate relations. In addition, it is worth mentioning that the overall shape of the derived luminosity functions is not very sensitive to the definition for the transition between the two phases.

Assuming that each SNR is a spherical shell of radius R, its luminosity in a given spectral band or emission line is:
\begin{equation} \label{eq:L}
\begin{aligned}
& L = 4\pi R(n,t)^2fL_{em}(v, \mu, n)\\
\end{aligned}
\end{equation}
where  $R$ is the SNR radius calculated by the first equation in the Eq. sets \ref{eq:R_sedov} or \ref{eq:snowplow}  depending on the evolutionary phase of the SNR, $f$ is the filling factor (i.e. the fraction of the shell area containing emitting gas), and $L_{em}(v, \mu, n)$ is the gas emissivity per unit area for a shock of a given velocity $v$, density $n$, and magnetic parameter $\mu$. The shock velocity is given by the second equation in the Eq. sets \ref{eq:R_sedov} or \ref{eq:snowplow},  while the density $n$ reflects the density distribution of the ISM in which the SNRs are embedded. We adopt the gas emissivity calculated by \citet{2008ApJS..178...20A}  for shocks of different conditions. This grid (of the physical parameters of the shock models \citealt{2008ApJS..178...20A}) covers velocities from $v=$ 100 $km\,s^{-1}$ to 1000 $km\,s^{-1}$, expanding in media with densities $n=$ 0.01 to 1000 \cmd,  and magnetic parameters $\rm \mu = B/\sqrt{n}$ between $ 10^{-4}\, \rm and\, 100 \, \mu \, G\, cm^{3/2} $.
Although the grids of \citet{2008ApJS..178...20A} include models for sub-solar, solar, and super-solar abundances, in this work we focus on solar abundance models, which are calculated for the widest range of pre-shock densities ($\rm n = 0.01 - 1000\, cm^{-3}$). We use the shock models with photoionizing precursor because they better represent the
observation of unresolved sources (as observed the extragalactic SNRs).   


\subsection{Calculation of luminosity functions} \label{LF_calculation}

\par
\begin{figure}
 \includegraphics[width=0.45\textwidth]{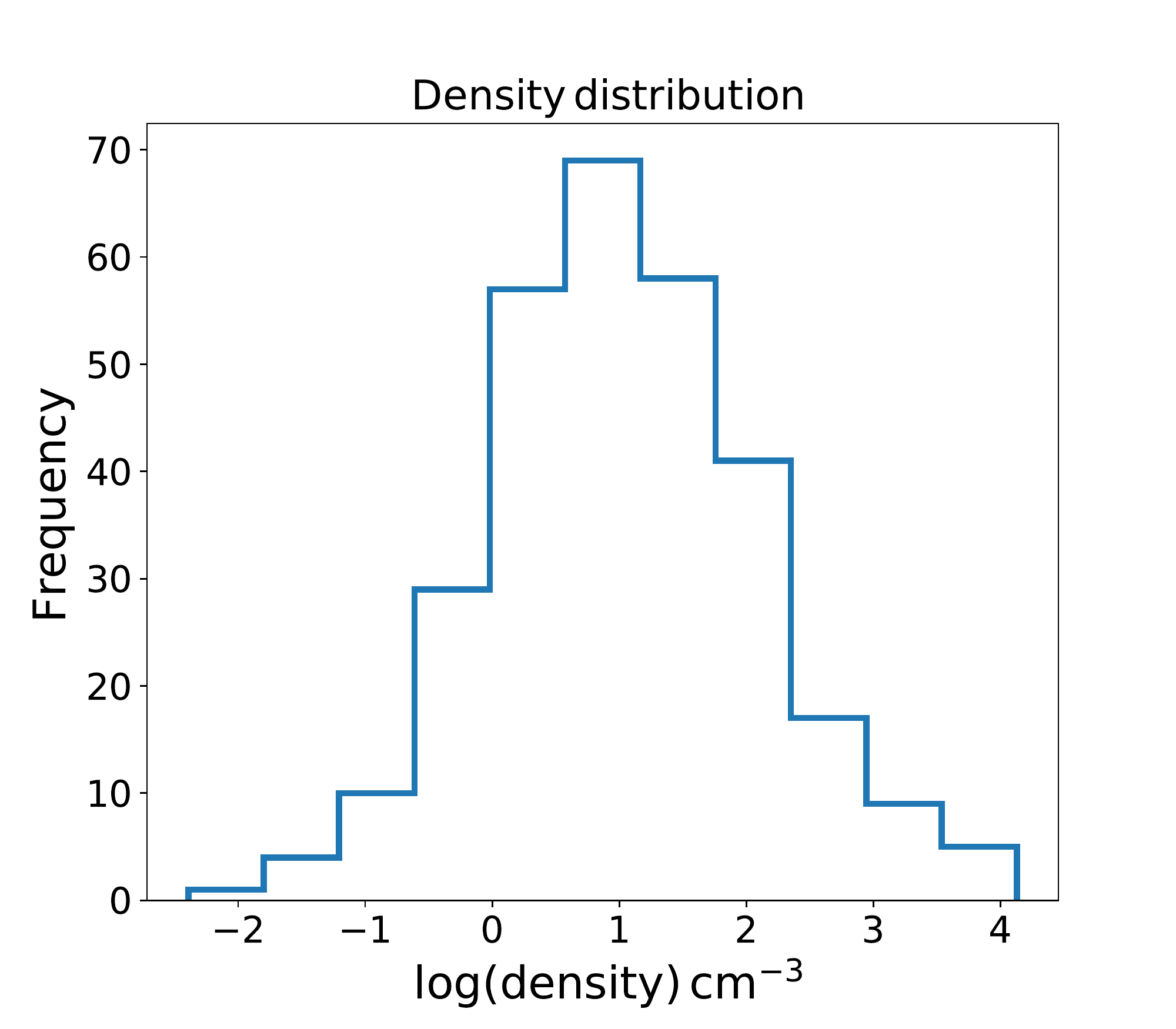}
  \caption{The baseline density distribution that we use in order to construct the theoretical LFs. It is a log-normal distribution with $\rm \mu(log(n/cm^{-3})) = 1.0$ and $\rm \sigma(log(n/cm^{-3})) = 1.0$.}
  \label{fig:density}
\end{figure}

 In order to calculate the luminosity distribution of an SNR population we assume: (a) a uniform age distribution between $500$ and $50000$ years and (b) a log-normal distribution of ISM density with mean $\rm \mu(log(n/cm^3)) = 1.0$ and standard deviation of $\rm \sigma(log(n/cm^3)) = 1.0$ (\autoref{fig:density}). The density distribution is a simplified assumption for the ISM,  which can be further improved and/or be adjusted depending on the application. 

\par By sampling the age and density from these independent distributions we can determine whether an SNR of a given age and density is in the adiabatic or the radiative phase from Eq. \ref{eq:tr}, and then calculate the corresponding radius, shock velocity, and luminosity using Eqs. \ref{eq:R_sedov}, \ref{eq:snowplow}, and \ref{eq:L}.
The gas emissivity is calculated by interpolating within the density and velocity grid of \citet{2008ApJS..178...20A}. Objects that have density or velocity that falls outside the grid are assigned the emissivity for the available models with the closest value for these parameters. 


The models of \citet{2008ApJS..178...20A} also depend on the shock magnetic parameter ($\mu$). However, since we do not have any information on the distribution of the magnetic parameters in SNRs, we perform our analysis for six different magnetic-parameter ranges: i) $\rm 10^{-4}-0.1\, \mu  G\, cm^{3/2}$; ii) $\rm 0.1-1\, \mu  G\, cm^{3/2}$;  iii) $\rm 1-3\, \mu  G\, cm^{3/2}$; iv) $\rm 3-4.5\, \mu  G\, cm^{3/2}$; v) $\rm 4.5-10\, \mu G\, cm^{3/2}$; vi) $\rm 10-100\, \mu  G\, cm^{3/2}$. This way we can assess the effect of the magnetic parameter on the resulting luminosity functions. 

\par In the following analysis we focus on the calculation of the \halpha and the joint \halpha - \sii luminosity functions of SNRs. However, following the above framework we can calculate the luminosity function in any spectral line or continuum band.

\par In order to compare the theoretical luminosity functions with observed samples of SNRs we have to apply the same selection criteria, as the ones  used in the observational studies.  The most typical criterion for the identification of SNRs is their \sii/\halpha > 0.4 ratio (\citealt{Mathewson&Clarke}). Although this can bias the SNR samples against slower velocity or older SNRs (\citealt{2020MNRAS.491..889K}), most SNR samples have been selected on this basis, so in the following analysis we only consider systems with \sii/\halpha luminosity ratios greater than 0.4.
For such models  and for each magnetic-field range, we construct the \halpha and the joint [\ion{S}{II}]-\halpha luminosity functions. These luminosity functions include the contribution of SNRs in the Sedov-Taylor as well as the radiative phase.

\section{Results} \label{results}

\subsection{\halpha luminosity function - Filling factor}
\begin{figure}
 \includegraphics[width=0.45\textwidth]{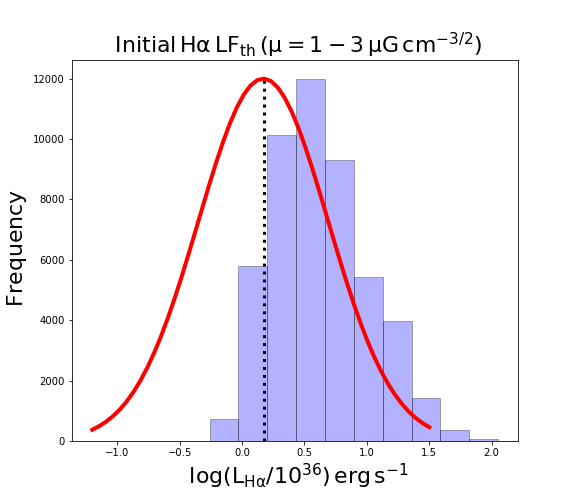}
 \caption{The blue histogram shows the model \halpha luminosity function for  magnetic parameters between $1$ and $3\, \mu G\, cm^{3/2}$ and filling factor $f = 1$. The red line is the incompleteness corrected observational \halpha luminosity function of SNRs in nearby galaxies  (\citealt{Kopsacheili2021}) rescaled arbitrarily along the y axis, and the dotted line shows its mode.}
 \label{fig:Ha_LF_wtht_ff}
\end{figure}

In \autoref{fig:Ha_LF_wtht_ff} (blue histogram) we show  the derived \halpha LF, for an SNR sample with an age and ambient ISM density distribution as described in $\S$ \ref{LF_calculation}, magnetic parameters from $1$ to $3\, \mu  G\, cm^{3/2}$, and filling factor $f=1$. Its shape is very similar to the observed luminosity function of SNRs in nearby galaxies (red line; \citealt{Kopsacheili2021}). However, the peak of the theoretical \halpha LF is at luminosities $\sim$0.3 dex higher than expected from the observations. This happens because in the calculation of the luminosities, we assumed a filling factor $f=1$ in Eq. \ref{eq:L}. Although it is difficult to measure directly the filling factor, especially for extragalactic SNRs,  simulations give values from 3\% to 23\% (e.g.  \citealt{2017ApJ...846...77S}). Indeed we find that we can match the observed and model LF for magnetic parameters  $1 - 3\, \mu  G\, cm^{3/2}$ (\autoref{fig:Ha_LF_wtht_ff}) by shifting the latter to lower luminosities by 0.66 dex. Similar offsets ($\sim $0.6 - 0.7 dex) are found for the model luminosity functions calculated for different magnetic parameters.  These correspond to filling factors of 25\% and 20\% respectively, in good agreement with the aforementioned simulations. In the following analysis, for the comparison with the observed data, we calculate the corresponding filling factor for each set of simulations so the mode of the observed and the model distribution match. We note that this shift along the luminosity axis does not affect the shape of the LF, which is the main subject of the following discussion.


\par
\begin{figure*}
           \hspace{-0.8cm}  \includegraphics[width=0.45\textwidth]{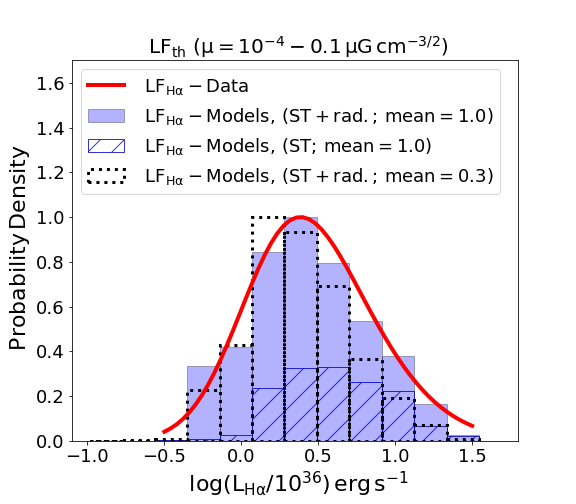}\hspace{0.8cm}
            \includegraphics[width=0.45\textwidth]{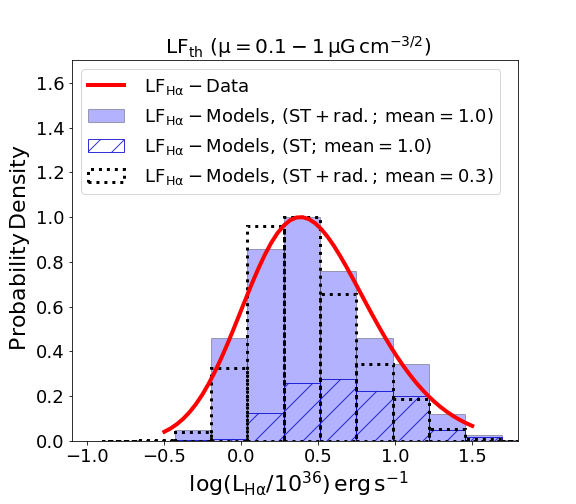}
            
            \hspace{-0.8cm} \includegraphics[width=0.45\textwidth]{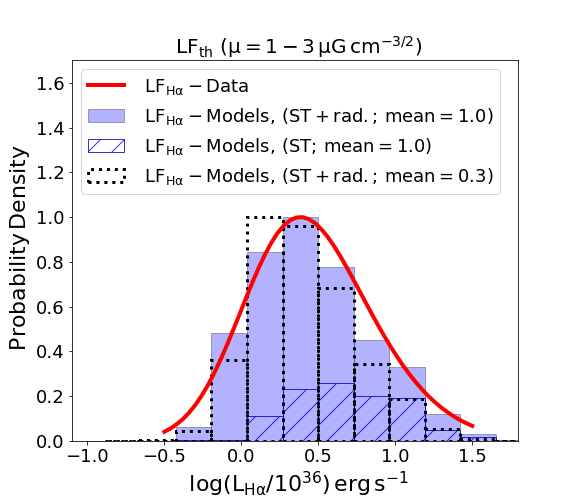}\hspace{0.8cm}
            \includegraphics[width=0.45\textwidth]{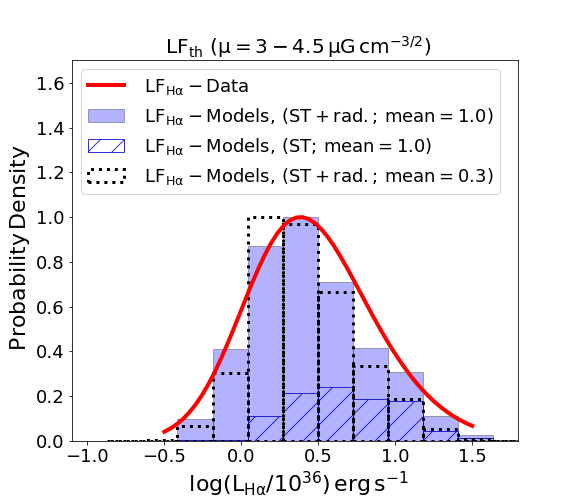}
           
           \hspace{-0.8cm}  \includegraphics[width=0.45\textwidth]{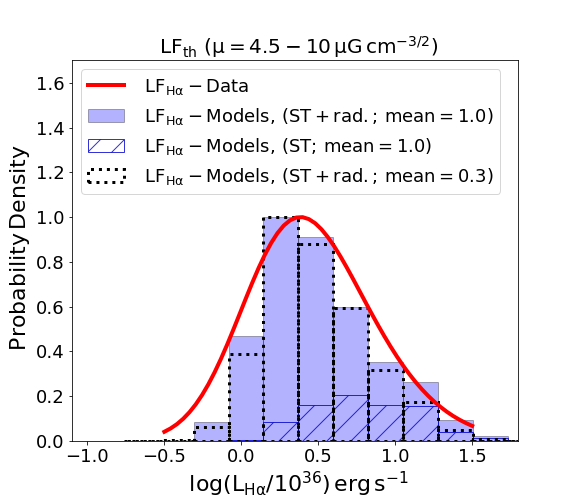}\hspace{0.8cm}
            \includegraphics[width=0.45\textwidth]{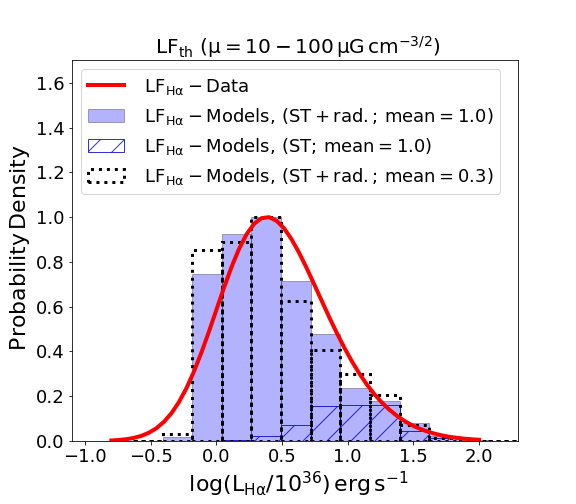}

        \caption[]
        {The blue histogram shows the  \halpha luminosity function resulting from our model for a density distribution with a mean of $\rm log(n/cm^{3})=1.0$  and standard deviation $\rm log(n/cm^{3})=1.0$. We only show SNRs with \sii/\halpha > 0.4. The hatched histogram shows the population of the simulated SNRs that are in Sedov-Taylor phase. For comparison we also show the model \halpha LF for a density distribution with a mean of $\rm log(n/cm^{3})=0.3$  and standard deviation of $\rm log(n/cm^{3})=1.0$ (dotted-line histogram). The red line shows the observed \halpha LF of SNRs of nearby galaxies from the work of \citet{Kopsacheili2021}. Different panels show the luminosity functions  for different magnetic parameters.}   
        \label{fig:Ha_LF}
    \end{figure*}

 The results of the \halpha LF including the filling factor correction are presented in \autoref{fig:Ha_LF} (blue histograms). The different panels give the resulting \halpha luminosity functions for the different regimes of magnetic parameters.  For reference we present also the LF of the simulated SNRs that are in  the Sedov-Taylor phase with hatched histograms. The dotted histogram shows the LF for the same model but for a less dense ISM (mean of $\rm log(n/cm^{-3}) = 0.3$). As can be seen, in all cases the theoretical \halpha luminosity functions closely resembles a Gaussian or a skewed Gaussian distribution that range from $\rm L \approx 0.3\times 10^{36}\rm erg\,s^{-1} $ to  $\rm L \approx 32 \times 10^{36}\rm erg\,s^{-1} $. 


\subsection{Joint [S {\small{II}}] - \halpha Luminosity Function}

In \autoref{fig:Ha_LF_1} (left-hand panels) we show the simulated SNR populations on the \halpha - \sii plane (blue circles) and the best fit relation of the form $\rm log(L_{[S\,II]}) = \alpha log(L_{H\alpha}) + b$ (blue line). The colorbar indicates the pre-shock density (darker colors correspond to higher density), and the size of the circles the shock velocity (larger circles indicate higher velocity). The joint \sii - \halpha LF is calculated along the best-fit line; i.e. we project the simulated SNRs on the best-fit line (blue line in \autoref{fig:Ha_LF_1}) and we examine their distribution along it. The joint \sii - \halpha luminosity functions for each magnetic parameter regime are presented in the right-hand panels of  \autoref{fig:Ha_LF_1}. Again, for reference, with the hatched histogram we present  the fraction of SNRs that are in the Sedov-Taylor phase. As in the case of the \halpha luminosity functions, we see that for all the magnetic parameters the theoretical \halpha luminosity functions are similar to a Gaussian or a skewed Gaussian.

\begin{figure*}
\hspace{-0.8cm}  
          \includegraphics[width=0.43\textwidth]{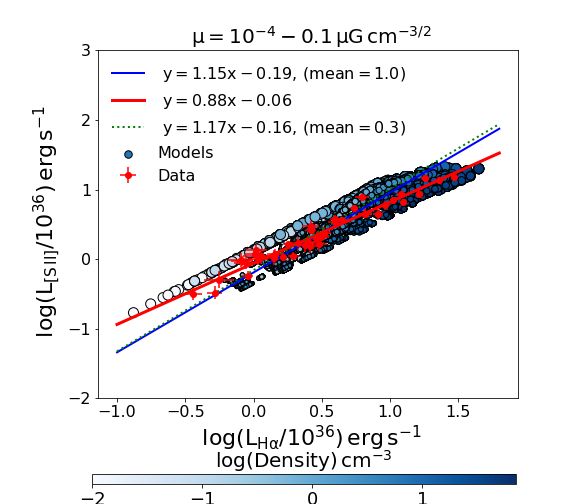}\hspace{0.8cm}
            \includegraphics[width=0.43\textwidth]{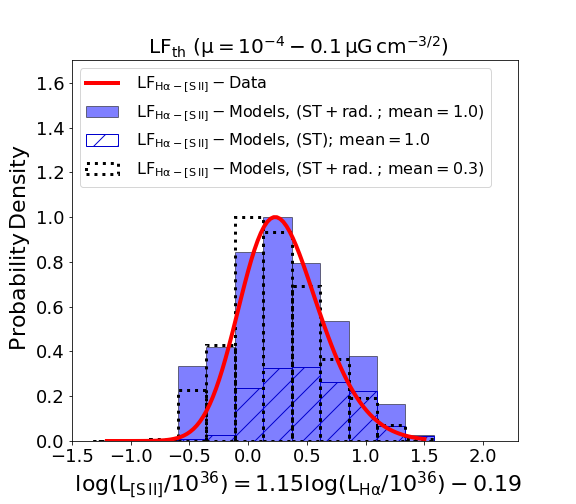}
            
         \hspace{-0.8cm}  \includegraphics[width=0.43\textwidth]{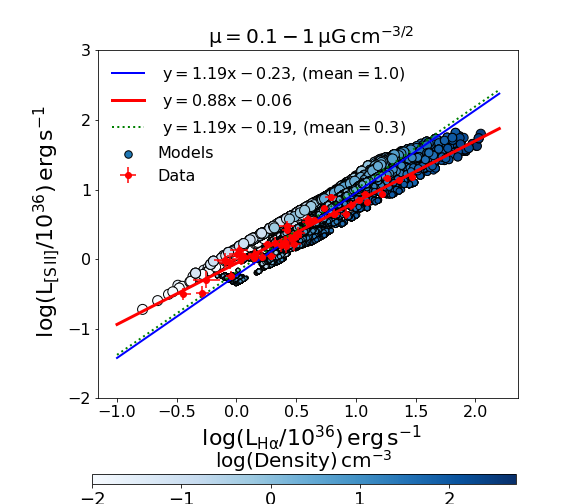}\hspace{0.8cm}
         \includegraphics[width=0.43\textwidth]{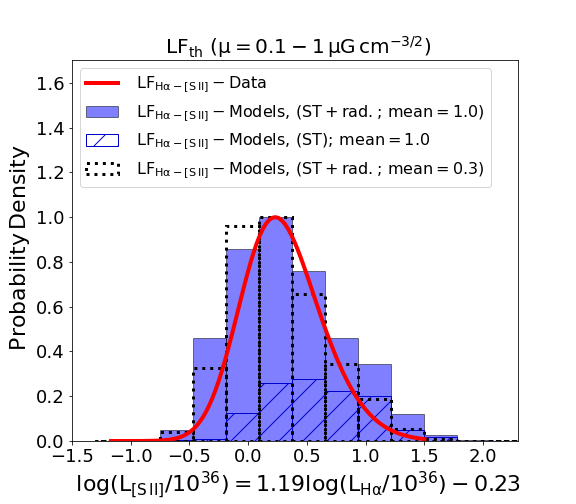}

         \hspace{-0.8cm}   \includegraphics[width=0.43\textwidth]{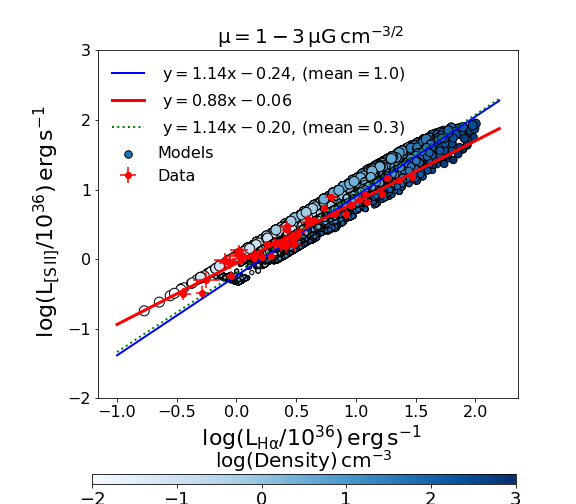}\hspace{0.8cm}
         \includegraphics[width=0.43\textwidth]{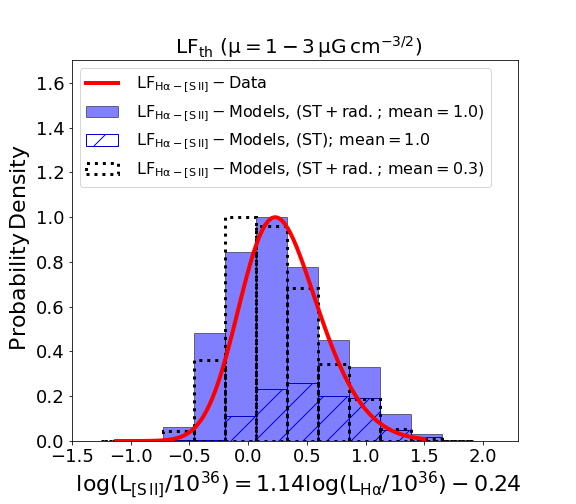}

        \caption[]
        {Left panels: the simulated SNRs on the \sii - \halpha plane (blue circles) for the model with  density distribution with a mean of $\rm log(n/cm^{3})=1.0$ and standard deviation $\rm log(n/cm^{3})=1.0$, and increasing magnetic parameters ($\rm \mu=10^{-4} - 3\, \mu G\, cm^{-3/2} $; top to bottom). The colorbar shows the pre-shock density (darker colors indicate higher density), while the size of the points indicate the shock velocity (larger circles indicate higher velocity). The blue line is the best fit to the $\rm L_{H\alpha} - L_{[S\, II]}$ relation of the model SNRs for density distribution with mean $\rm log(n/cm^{3})=1.0$, while the green dotted line is the best-fit line for density distribution with mean $\rm log(n/cm^{3})=0.3$ (the uncertainties in the slope and intercept of the lines are $\le 1\%$). The red points show the SNRs detected in  nearby galaxies (\citealt{Kopsacheili2021}) and the red line their best-fit $\rm L_{H\alpha} - L_{[S\, II]}$ relation. Right panels: the blue histogram shows the model joint \sii-\halpha luminosity function (LF) along the best-fit $\rm L_{H\alpha} - L_{[S\, II]}$ relation shown in the abscissa (blue line). The hatched histogram shows the population of the simulated SNRs that are in Sedov-Taylor phase. The dotted-line histogram shows the theoretical \sii- \halpha LF of the overall SNR population (in the Sedov-Taylor and the radiative phase) for a density distribution with $\rm log(n/cm^{3})=0.3$ using the same filling factor as for the baseline density distribution. For comparison we also show with the red line the \sii-\halpha LF of SNRs in nearby galaxies (\citealt{Kopsacheili2021}) along the red line of the left-hand panel figures.  }
        \label{fig:Ha_LF_1}
    \end{figure*}

\begin{figure*}
\ContinuedFloat
\hspace{-0.8cm}  
          \includegraphics[width=0.43\textwidth]{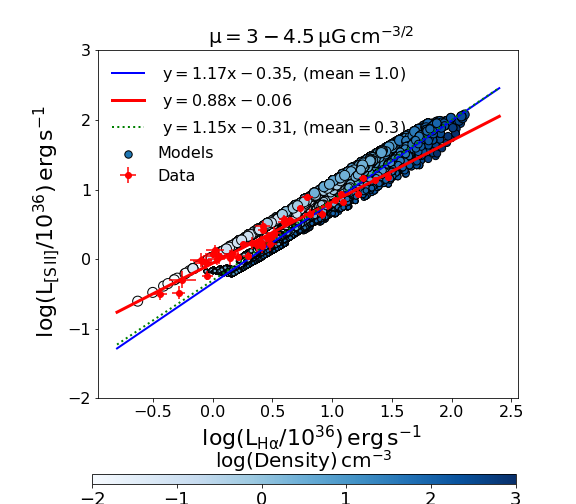}\hspace{0.8cm}
            \includegraphics[width=0.43\textwidth]{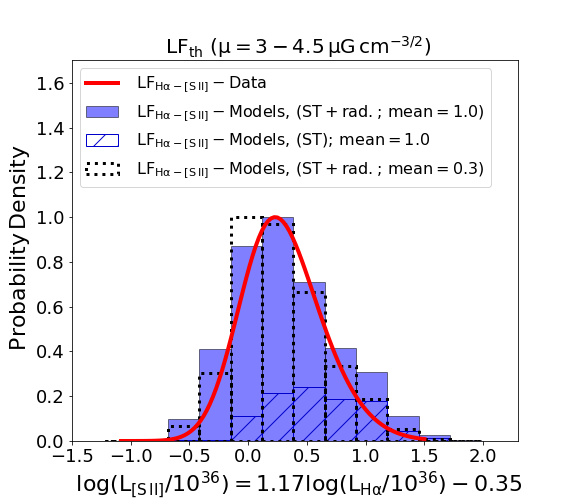}
            
         \hspace{-0.8cm}  \includegraphics[width=0.43\textwidth]{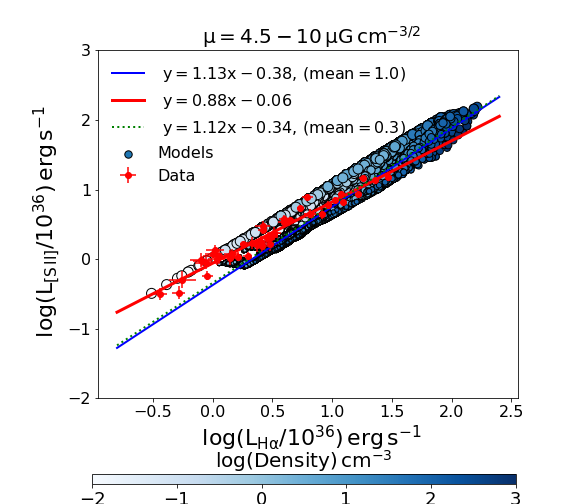}\hspace{0.8cm}
         \includegraphics[width=0.43\textwidth]{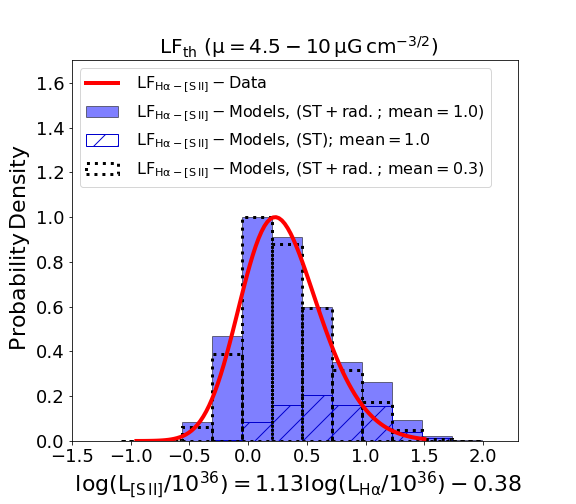}

         \hspace{-0.8cm}   \includegraphics[width=0.43\textwidth]{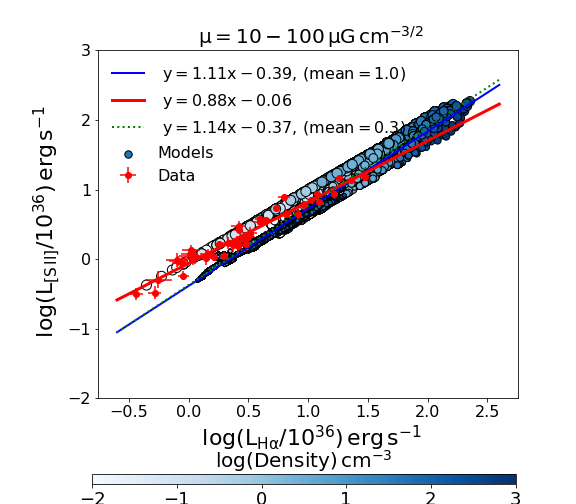}\hspace{0.8cm}
         \includegraphics[width=0.43\textwidth]{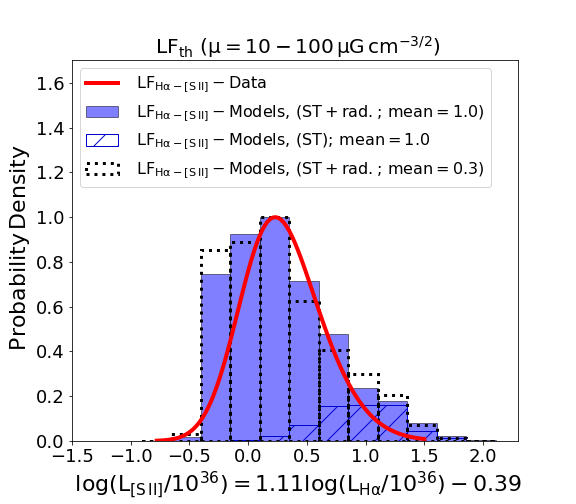}

        \caption[]
       {{\it{\bf{continued.}}} Same as fig. 4 but for magnetic parameters $\rm \mu=3 - 100\, \mu G\, cm^{-3/2} $.}  
       
       \label{fig:Ha_LF_1}
    \end{figure*}

\section{Discussion} \label{discussion}
In this work, we present the first attempt in building a population synthesis model for the luminosity distribution of SNRs. Based on a baseline SNR evolution model, a library of shock excitation models, and assumptions on the density and age distribution of SNR populations, we can predict the luminosity distribution of SNRs in different spectral lines. Next we discuss the qualitative comparison of these models, with the observed \halpha and (\halpha, \sii) luminosity distribution of SNRs.

\subsection{Luminosity Functions - Comparison with data} \label{data}
In \autoref{fig:Ha_LF} and in the right-hand panels of  \autoref{fig:Ha_LF_1}, we see the theoretical \halpha and the joint \sii- \halpha  luminosity functions respectively (blue and hatched histograms). The blue histogram shows the overall population of SNRs with \sii/\halpha > 0.4 while the hatched histogram indicates the SNRs that are in Sedov-Taylor phase, according to the definition that we have adopted.

\par
\begin{figure}
\includegraphics[width=0.5\textwidth]{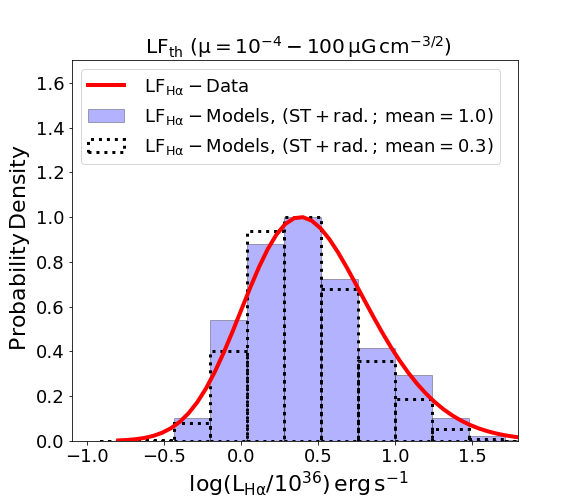}
\caption[]
        {The blue histogram shows the  \halpha luminosity function resulting from our model for the full magnetic-parameter range ($\rm \mu=10^{-4} - 100\, \mu G\, cm^{-3/2} $), and for a density distribution with a mean of $\rm log(n/cm^{3})=1.0$  and standard deviation $\rm log(n/cm^{3})=1.0$. We only show SNRs with \sii/\halpha > 0.4. For comparison we also show the model \halpha LF for a density distribution with a mean of $\rm log(n/cm^{3})=0.3$  and standard deviation of $\rm log(n/cm^{3})=1.0$ (dotted-line histogram). The red line shows the observed \halpha LF of SNRs of nearby galaxies from the work of \citet{Kopsacheili2021}.}
\label{fig:LF_Ha_all}
\end{figure}

\begin{figure*}
\hspace{-0.8cm}  
          \includegraphics[width=0.43\textwidth]{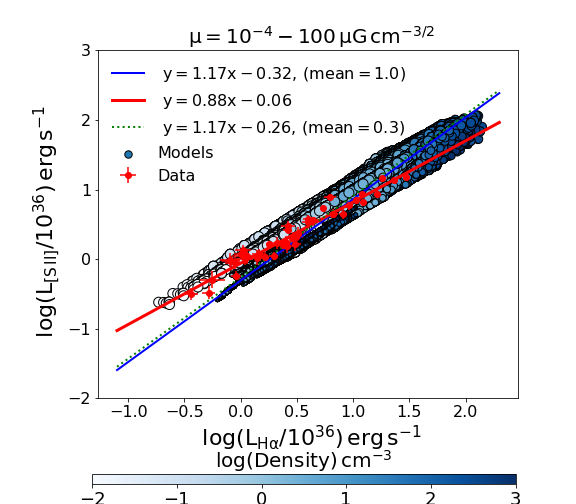}\hspace{0.8cm}
            \includegraphics[width=0.43\textwidth]{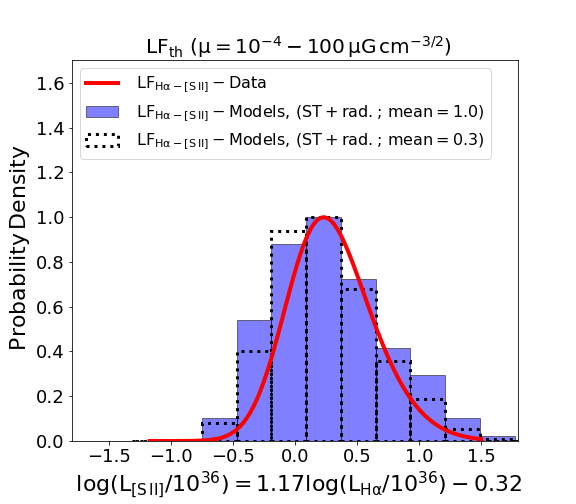}
            \caption[]
        {Left panel: the simulated SNRs on the \sii - \halpha plane (blue circles) for the full magnetic-parameter range ($\rm \mu=10^{-4} - 100\, \mu G\, cm^{-3/2} $), and for the model with  density distribution with a mean of $\rm log(n/cm^{3})=1.0$ and standard deviation $\rm log(n/cm^{3})=1.0$). The colorbar shows the pre-shock density (darker colors indicate higher density), while the size of the points indicate the shock velocity (larger circles indicate higher velocity). The blue line is the best fit to the $\rm L_{H\alpha} - L_{[S\, II]}$ relation of the model SNRs for density distribution with mean $\rm log(n/cm^{3})=1.0$, while the green dotted line is the best-fit line for density distribution with mean $\rm log(n/cm^{3})=0.3$. The red points show the SNRs detected in  nearby galaxies (\citealt{Kopsacheili2021}) and the red line their best-fit $\rm L_{H\alpha} - L_{[S\, II]}$ relation. Right panel: The blue histogram shows the model joint \sii-\halpha luminosity function (LF) along the best-fit $\rm L_{H\alpha} - L_{[S\, II]}$ relation shown in the abscissa (blue line). The dotted-line histogram shows the theoretical \sii- \halpha LF of the overall SNR population (in the Sedov-Taylor and the radiative phase) for a density distribution with $\rm log(n/cm^{3})=0.3$ using the same filling factor as for the baseline density distribution. For comparison we also show with the red line the \sii-\halpha LF of SNRs in nearby galaxies (\citealt{Kopsacheili2021}) along the red line of the left-hand panel figures. } 
       \label{fig:LF_2d_all}
\end{figure*}

 In \autoref{fig:Ha_LF} and in the right-hand panels of  \autoref{fig:Ha_LF_1}, we compare the theoretical luminosity functions with those measured in 4 nearby galaxies (red line; \citealt{Kopsacheili2021}). We see that there is a very good agreement between the theoretical and observational  luminosity functions more or less for all the magnetic parameters.  In figures \ref{fig:LF_Ha_all} and \ref{fig:LF_2d_all}  we compare the observed luminosity function (red line) with the theoretical ones (\halpha and \sii-\halpha respectively) for the full magnetic-parameter range. We see an even better agreement between the theoretical and the observed luminosity functions, which is an indication that the SNR population is characterized by a broad distribution of magnetic parameters. In some of these cases,  and especially for the joint LF (figures \ref{fig:Ha_LF_1} and \ref{fig:LF_2d_all}) we see that the theoretical luminosities have a  broader luminosity range than the data, extending to lower luminosities. This happens because the observed population of candidate SNRs presented in this study (\citealt{Kopsacheili2021}) were selected such as their \halpha luminosity is 3$\rm \sigma$ above the background and their \sii/\halpha ratio 3$\rm \sigma$  above the 0.4 threshold. These limits exclude especially sources with lower luminosity. This becomes particularly important for the joint \sii - \halpha LF where the selection is driven by the weaker \sii lines. Nonetheless, our results indicate that the model luminosity functions cover very similar luminosity range as the observed ones. Furthermore, based on our filling factor scaling, the simulated luminosity functions trace the vast majority of our observed SNR populations.

\par
\begin{figure*}
\includegraphics[width=1.0\textwidth]{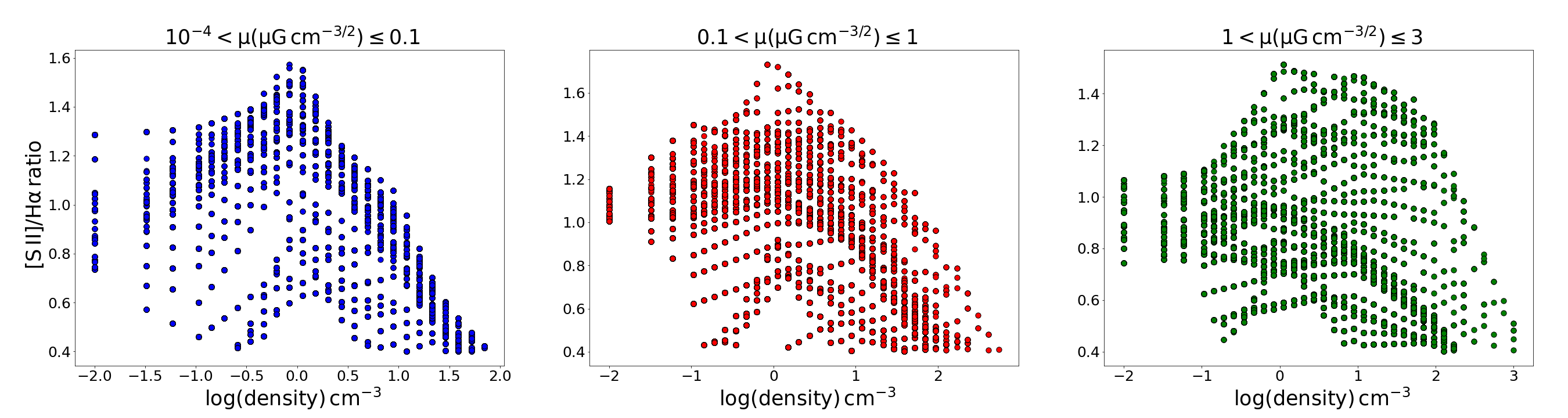}
\caption[]
        {The \sii/\halpha ratio of the simulated SNRs that exceed the \sii/\halpha = 0.4 threshold, as a function of the pre-shock density for magnetic parameters between $10^{-4}-3\, \mu G\, cm^{3/2}$.}
\label{fig:den_rat}
\end{figure*}

 In the left-hand panels of figures \ref{fig:Ha_LF_1} and \ref{fig:LF_2d_all} we show the simulated SNRs (blue circles) and the observed SNRs (red points) on the \sii - \halpha plane. In the simulated SNRs, the lower limit of the \sii luminosity for a given \halpha luminosity, is determined by the \sii/\halpha $>$ 0.4 ratio. This selection criterion has been applied in our models in order to be able to compare our results with the observed SNR populations that have been selected based on this criterion. The upper limit is determined by the shock velocity and the pre-shock density of the models of \citet{2008ApJS..178...20A}. In general, as we see, there is a trend for the more luminous observed SNRs to present lower \sii luminosities than the simulated ones. More specifically, the more luminous observed SNRs correspond to SNRs with higher pre-shock density (i.e. darker blue circles). In order to explore this behavior, in \autoref{fig:den_rat} we examine the shock excitation (i.e. \sii/\halpha ratio) of the simulated SNRs as a function of the pre-shock density. We only show models for magnetic parameters  $\rm 10^{-4}-3\, \mu G\, cm^{3/2}$, but the behavior is similar for other magnetic parameters.  In each case, different points for the same density correspond to different shock velocities, generally (but not always) increasing from bottom to top. The scatter of the points becomes larger because the magnetic parameter range becomes wider in the right-hand plots. In all cases there is a  density above which the \sii/\halpha ratio decreases. This could be the result of the shock excitation becoming lower for higher density because of the increasing effect of the collisional de-excitation. This trend is also seen in the observed SNR populations  (e.g. red points in \autoref{fig:Ha_LF_1}) where the SNR distribution on the \halpha - \sii plane presents a sub-linearity (Fig. 8 and 10 in \citealt{Kopsacheili2021}). In the data this trend is magnified since SNRs in denser environments are more likely to be embedded in \ion{H}{II} regions. This is reflected in the shallower $\rm log(L_{[S\, II]}) - log(L_{H\alpha})$ relations fitted to the observed SNR populations with the red line in \autoref{fig:Ha_LF_1}.


\subsection{Effect of density distribution}
The evolution of SNRs and their luminosity are highly sensitive on the density ($n$) of the medium in which they expand (Eqs. \ref{eq:R_sedov} and \ref{eq:snowplow}). Thus, we expect to have higher luminosity SNRs in regions with higher pre-shock densities. This behavior is predicted by detailed SNR evolution models (e.g. \citealt{2005MNRAS.360...76A}). It is also seen in observational data, where there is a lack of bright SNRs in regions with low \halpha background indicating low density environments (\citealt{Kopsacheili2021}). The same behaviour is observed in our models. As we see in  \autoref{fig:Ha_LF_1} (left-hand panels) the simulated SNRs with higher pre-shock density (darker blue circles) present higher luminosities. 

\par  In order to assess the effect of the assumed density distribution on the resulting luminosity functions, we explore models with different density distributions. In Figures \ref{fig:Ha_LF} and \ref{fig:Ha_LF_1} the dotted-line histograms and the dotted lines show the theoretical luminosity functions  and the best fit line on the \sii - \halpha plane respectively for log-normal distributions but with a mean value $\rm log(n/cm^{-3}) = 0.3 $  instead of $\rm log(n/cm^{-3}) = 1 $. 

The shape of the  luminosity functions is now different. From \autoref{fig:Ha_LF} we see that in general the low-density model results in rather narrower \halpha luminosity functions showing particularly a deficit of high luminosity objects. This trend holds also for the joint \sii-\halpha luminosity function. 

The fact that different density distributions result in different luminosity functions indicates that these models can be used in order to determine the density distribution of the SNR environment based on their luminosity functions, using representative and bias-free samples of SNRs.


\subsection{Effect of velocity distribution}
As the SNR evolves, more material accumulates behind the shock decreasing its velocity (e.g. \citealt{2017hsn..book.1981R}). With the deceleration of the shock, the excitation of the shocked material also decreases resulting in lower luminosities. Hence, we expect that higher luminosity SNRs are associated with higher shock velocities, which agrees with the model predictions. 

\begin{figure*}
\includegraphics[width=1.0\textwidth]{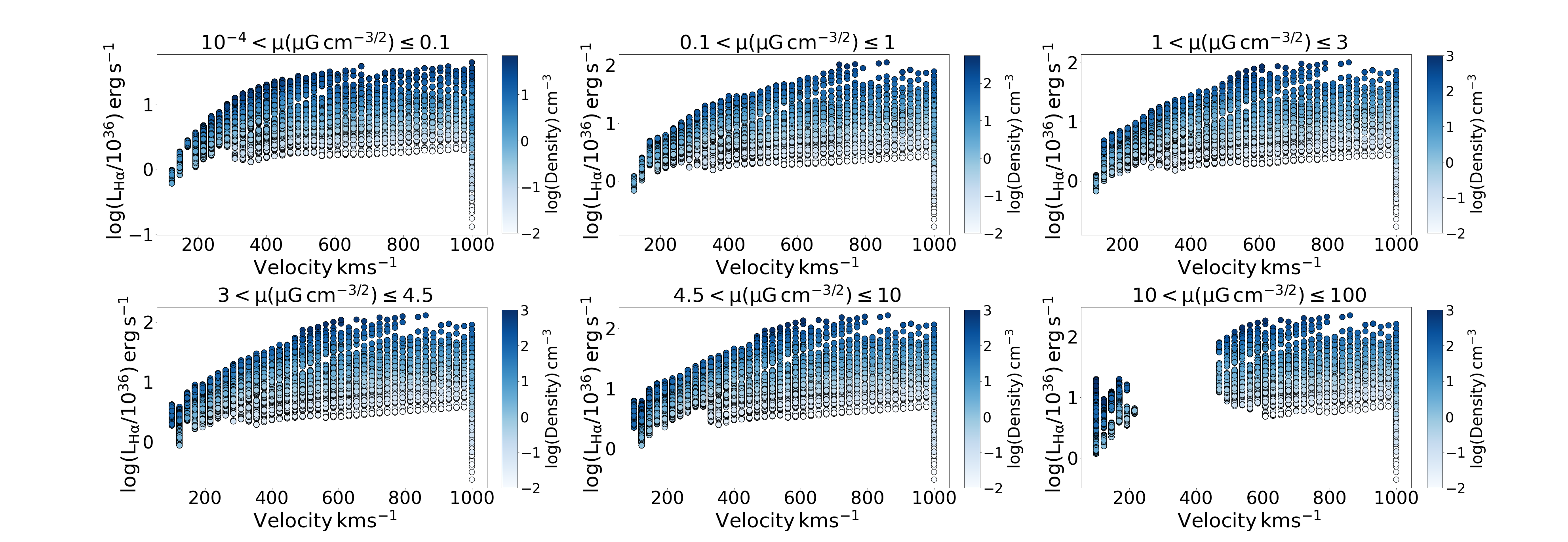}

\hspace{0.5cm}
\includegraphics[width=1.0\textwidth]{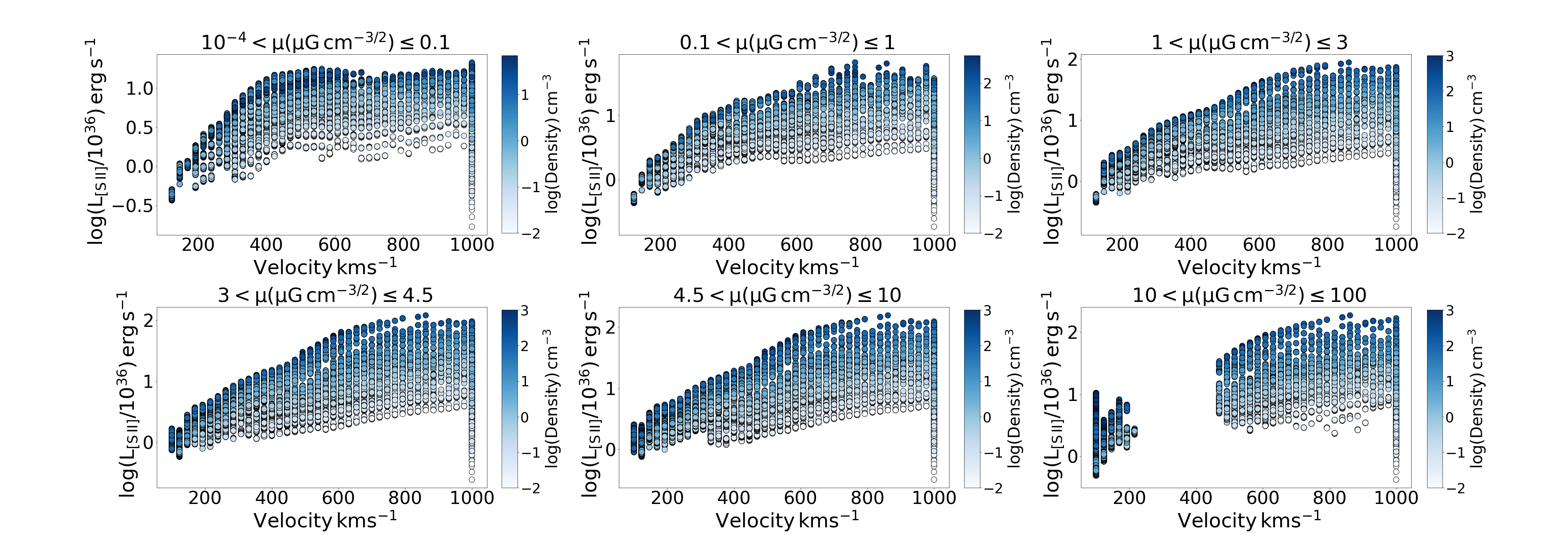}
\caption[]
        {The \halpha and \sii simulated luminosities as function of the shock velocity. The blue colorbar indicates the pre-shock density. Darker colors correspond to higher densities}
\label{fig:vel_Ha_SII}
\end{figure*}

 This can be seen in \autoref{fig:Ha_LF_1} where larger circles indicate higher velocities, and more directly in \autoref{fig:vel_Ha_SII}. The latter shows  the \halpha and \sii luminosities as a function of the shock velocity. The points are color-coded according to the pre-shock density (darker colors indicate higher density). We note that the models accumulated at the velocity of  1000 $km\,s^{-1}$ are those with velocities $\ge 1000\, km\, s^{-1}$  that falls outside the grid published by \citet{2008ApJS..178...20A} and as  described in \ref{LF_calculation} are assigned the
emissivity of the available models with the closest values to their physical parameters. These consist only $\sim$ 3\% of the total number of models and thus they do not affect significantly the shape of the luminosity function. 
\par However, we also see models that despite their high velocity \\ present low luminosities. These are models with low pre-shock densities (lighter blue circles) in  figures \ref{fig:Ha_LF_1} and \ref{fig:vel_Ha_SII}. These models represent SNRs that expand in low-density environments, where shock excitation is not very efficient resulting in lower luminosities.

\subsection{Effect of magnetic parameters} \label{mag_par}
\par The shock models that we used (\citealt{2008ApJS..178...20A})  to construct the theoretical  luminosity functions are given for a wide range of magnetic parameters ($\rm \mu  = 10^{-4} - 100 \, \mu \, G\, cm^{3/2}$). Since we do not have any information on the magnetic parameters of observed SNRs, we calculated the luminosity functions for six different regimes of  magnetic parameters. As we see in  Figures \ref{fig:Ha_LF} and \ref{fig:Ha_LF_1}, the theoretical \halpha and the joint \sii - \halpha luminosity functions are quite similar to the luminosity functions of SNRs observed in nearby galaxies (\citealt{Kopsacheili2021}). However, for magnetic parameters higher than $\rm 10 \, \mu  G\, cm^{3/2}$, we see a slight excess of SNRs in lower luminosities. 
This happens because stronger magnetic fields (indicated by the higher magnetic parameters since the densities are more or less the same for all the cases) confine the post-shock region, limiting the  further increase of the post-shock density that would result in more luminous shocks (e.g. fig. 5 in \citealt{2008ApJS..178...20A}). This phenomenon is more evident in low velocities (100 $km\,s^{-1}$) where the magnetic field has a more important effect.

\subsection{Implications and Limitations}
The presented framework for building model luminosity functions of SNRs provides a useful tool for exploring the  properties of SNRs in the context of their evolution. Although the model presented here can be considered as a "toy" model, it lays the foundation for a population synthesis model for SNRs. 
Such a model can be used to explore the effect of different underlying distributions for physical parameters, such as density, shock velocity, and magnetic parameter on the observed luminosity distributions. When appropriate observational data are available, one also could use these models to infer these distributions.

\par However, we recognize that there are limitations in the current state of models, that  prevent us from making reliable inference of the physical parameters. These limitations arise from several factors the most important of which are: i) The grid used to obtain the luminosity of SNRs as a function of their shock velocity, density, and magnetic parameter, covers a rather narrow range of densities and shock velocities. Extending it to lower shock velocities will allow us to model older SNRs or those expanding in dense environments, while higher shock velocities will be useful for SNRs expanding in low density environments. This way we will avoid boundary effects like those mentioned in $\S$ \ref{LF_calculation}. Similarly, the density grid does not account for SNRs expanding in very low or very high density environments; ii) Another limitation of this toy model is the basic treatment of SNR evolution (i.e. the equations of the different evolutionary phases). Prescriptions and fitted formulae based on more realistic, hydrodynamic models can be more representative of the real evolutionary processes in SNRs; iii) The unknown filling factor prevents us from accurately adjusting the luminosity range of the calculated LF. Although this does not affect the shape of the model luminosity functions it does not allow us to take advantage of the full diagnostic power of this model; iv) The treatment of the SNRs as perfectly spherical shells can also introduce uncertainties in the comparison with real data.

\par These limitations will be overcome once detailed maps of Galactic and Magellanic Clouds SNRs in different spectral lines and continuum bands (e.g. from IFU data) become available. Such data will show the departure of the SNR shape from the assumed spherical shell, variations of the filling factor and excitation within the SNR and as a function of its age, and can be used to test SNR evolution prescriptions in future versions of the models.

\par In our simulated SNRs there is significant degeneracy of the observable parameters as a function of the SNR physical parameters: i.e. two SNRs with the same shock velocity, density, and magnetic parameter may have different \halpha and \sii luminosity. These degeneracies can be reduced to some degree by including in the models additional observable quantities (e.g. other spectral lines, or continuum bands). 

\par This will help using these models as a diagnostic tool for constraining the underlying physical parameters and processes of the SNRs once some of the above limitations are remedied. Furthermore, by integrating over the physical parameters of the SNR population, we can estimate their integrated luminosity output in different bands or their total momentum/mechanical energy output. This is particularly useful for constraining feedback processes in different galactic environments (e.g. by modeling SNR populations observed in different galaxies) or by estimating their contributions in the measured emission in different spectral lines.

\section{Conclusions} \label{conlusions}
In this work, we present a first theoretical model for the populations of supernova remnants. We use shock models, that give the luminosity of emission lines as a function of the physical parameters of the shock (shock velocity, pre-shock density and magnetic parameter). Considering a flat distribution for the age of SNRs and a log-normal distribution for their pre-shock density, we construct theoretical \halpha and  joint \sii - \halpha luminosity functions. We find remarkable agreement between the predicted luminosity functions with the  luminosity functions of the observed SNR populations in nearby galaxies, after correcting the models for a filling factor. Also, remarkably the estimated filling factor agrees very well with the estimated one from detailed SNR simulations. 

\par Although these models are far from being considered as complete, they provide the first step in the developing a framework for SNR population synthesis models. Such models will be particularly useful for exploiting the potential of future high resolution IFU observations and for modeling.

\section*{Acknowledgements}

We thank the anonymous referee for a thorough review and helpful comments that helped
to improve the clarity of the paper, We acknowledge funding from
the European Research Council under the European Union’s Seventh
Framework Programme (FP/2007-2013)/ERC Grant Agreement n.
617001. This project has received funding from the European Union’s
Horizon 2020 research and innovation programme under the Marie
Sklodowska-Curie RISE action, grant agreement No 691164 (ASTROSTAT). We also acknowledge support from the European Research Council under the European Union’s Horizon 2020 research
and innovation program, under grant agreement No 771282. IL acknowledges support by Greece and the European Union (European Social Fund - ESF) through the Operational Programme "Human Resources Development, Education
and Lifelong Learning" in the context of the project "Reinforcement of Postodoctoral Researchers-2nd Cycle" (MIS-5033021), implemented by the State Scholarships Foundation (IKY). 

\section*{DATA AVAILABILITY}

The data that have been used for comparison with the presented models are from the work of \citet{Kopsacheili2021}.
The raw data are available from the NOAO  data archives.






\twocolumn




\appendix


\bsp	
\label{lastpage}
\end{document}